\documentclass[a4paper,11pt]{article}
\usepackage{jinstpub} 

\usepackage{graphicx}
\usepackage{hyperref}
\usepackage{float} 
\usepackage{xcolor}
\usepackage{amsmath}

\usepackage{siunitx}
\DeclareSIUnit{\electronvolt}{eV}

\usepackage{amssymb} 

\usepackage[T1]{fontenc}
\usepackage[english]{babel}
\usepackage{subcaption}
\usepackage{textcomp,gensymb}
\usepackage{pifont} 
\usepackage{orcidlink} 

\title{A dump leakage calorimeter to measure the flux of high-energy electrons and photons}

\author[a,b]{Antonios Athanassiadis\orcidlink{0009-0008-9963-0024}}
\author[a]{Ties Behnke\orcidlink{0000-0001-5984-6156}}
\author[a,1]{Jonas Björklund Svensson\orcidlink{0000-0001-7161-9714}\note{Current address: Lund University, 22100 Lund, Sweden}}
\author[a,2]{Oleksandr Borysov\orcidlink{0000-0002-5384-611X}\note{Current address: Weizmann Institute of Science, 7610001 Rehovot, Israel}}
\author[a,2]{Maryna Borysova\orcidlink{0000-0002-4481-5872}}
\author[a]{Lewis Boulton}
\author[a,3]{Sarawit Chindaratchakul\orcidlink{0009-0006-4910-4814}\note{Current address: Mahidol University, 10400 Bangkok, Thailand}}
\author[a,b]{Beate Heinemann\orcidlink{0000-0002-1673-7926}}
\author[a]{Louis Helary\orcidlink{0000-0001-7891-8354}}
\author[a]{Ruth Jacobs\orcidlink{0000-0001-5446-5901}}
\author[a]{Advait Kanekar}
\author[a]{Jenny List\orcidlink{0000-0002-0626-3093}}
\author[a]{Tianyun Long\orcidlink{0000-0002-9357-1894}}
\author[a,4]{Tanguy Marsault\orcidlink{0009-0004-8467-9424}\note{Current address: CEA Paris-Saclay, 91190 Saclay, France}
}
\author[a,b,5]{Felipe Peña\orcidlink{0000-0001-6755-513X}\note{Current address: Ludwig-Maximilians-Universität München, 85748 Garching, Germany}}
\author[a]{Stefan Schmitt}
\author[a,6]{Sarah Schröder\orcidlink{0000-0002-0649-9995}\note{Current address: Lawrence Berkeley National Laboratory, Berkeley, CA 94720, USA}}
\author[a,c,7]{Ivo Schulthess\orcidlink{0000-0002-5621-2462}\note{Corresponding author}}
\author[a]{Stephan Wesch\orcidlink{0000-0002-3789-8884}}
\author[a,d]{Matthew Wing\orcidlink{0000-0002-6319-4135}}
\author[a]{Jonathan Wood\orcidlink{0000-0003-4413-7044}}

\affiliation[a]{Deutsches Elektronen-Synchrotron DESY, 22603 Hamburg, Germany}
\affiliation[b]{Universität Hamburg, 22761 Hamburg, Germany}
\affiliation[c]{Institute for Particle Physics and Astrophysics, ETH Zurich, 8093 Zurich, Switzerland}
\affiliation[d]{University College London, London WC1E 6BT, United Kingdom}

\emailAdd{ivo.schulthess@desy.de}

\abstract{We developed a novel apparatus based on a lead-glass calorimeter that can measure the flux of high-energy electrons or photons. Our detector uses the electromagnetic shower leakage from the beam dump, where the particles are disposed of at the beamline's end. A prototype of such a calorimeter was set up at the FLASHForward experiment at DESY. We show that it can measure the electron bunch charge with a typical precision on the order of 10\% and an accuracy at the few-percent level. Additionally, it is capable of determining the beam's position with a precision on the order of tens of micrometers. Finally, we demonstrate applicability to high-energy photons. }

\arxivnumber{2508.17991} 

\notoc

\begin{document}

\maketitle
\flushbottom

\section{Introduction}\label{sec:intro}

Accurately measuring high-flux, high-energy photons is a general experimental challenge. Such photon beams can originate from future experiments and facilities, such as the LUXE experiment that is planned at the Deutsches Elektronen-Synchrotron DESY in Hamburg, Germany~\cite{abramowicz_conceptual_2021, luxecollaboration_technical_2024}. LUXE aims to test quantum electrodynamics in the strong-field regime. The interaction of electron bunches from the European XFEL~\cite{abela_xfel_2006} with a high-intensity laser will produce pulses of ${\mathcal{O}\left(10^9\right)}$ photons with an energy spectrum of up to a few GeV. Similar photon sources could also be included in future circular or linear particle collider facilities such as the {FCC-ee} or LCF/CLIC/ILC, respectively~\cite{schulthess_new_2025}. These photon beams serve as key diagnostic tools for studying the strong-field quantum electrodynamics interaction and also offer promising opportunities as a probe in fixed-target experiments searching for new physics. 

The measurement of high-energy photon fluxes is also relevant for fixed-target or beam-dump experiments searching for new phenomena beyond the standard model of particle physics. This is motivated by the absence of new physics signals at high-energy colliders and the theoretical expectation that weakly interacting particles, potential candidates for dark matter or mediators of hidden sectors, may be more effectively probed in high-intensity, low-background environments~\cite{beacham_physics_2020}. In these experiments, high-energy particles are dumped onto a solid target, where new long-lived particles can be produced and reconstructed via their decay products. Knowing the flux and properties of the beam impinging onto the target is critical for determining production rates and exclusion limits. Various options exist to measure the flux of charged particles~\cite{lensch_comparison_2023}. However, for high-energy photon beams, non-invasive flux measurements remain particularly challenging, motivating the approach presented in this work.

Current diagnostic systems for high-flux, high-energy photon beams typically use a target to partially convert the photon beam into electron-positron pairs, which can then be measured using a magnetic dipole spectrometer~\cite{fleck_conceptual_2020, cavanagh_experimental_2023}. However, this method has the disadvantage of disturbing the photon beam.\footnote{Furthermore, it requires a dipole magnet, a carefully selected converter target which has to be stable during operation, and a sizable particle detector that covers the spatial spread after the magnet. } 

In this paper, we describe a method for measuring the photon-beam flux without interfering with the beam during its transport. Our novel approach uses a lead-glass calorimeter to detect the electromagnetic shower leakage emerging from the beam dump, where the photons are ultimately absorbed. Although the beam is terminated at the dump, our measurement is non-invasive with respect to the beam's propagation and interaction up to that point. The shower leakage, which is the part of the electromagnetic shower that is not contained in the dump, is known to correlate with the photon beam flux~\cite{luxecollaboration_technical_2024}. This technique allows for flux monitoring without introducing any upstream disturbance to the beam or requiring additional components in the beamline.

Commissioning and testing of this detector is difficult because, to date, there is no running high-flux high-energy pulsed photon source that matches the requirements of the diagnostic system. To test the dump leakage calorimeter in such an environment, we exploited the similarities of the electromagnetic shower development in beam dumps for both photons and electrons. Specifically, we set up a prototype at the dump of the FLASHForward beam-driven plasma-wakefield experiment at DESY, where we could test the detector parasitically to the FLASHForward measurement campaigns~\cite{darcy_flashforward_2019}.

\section{Setup}\label{sec:setup}

\begin{figure}
    \centering
    \includegraphics[width=\linewidth]{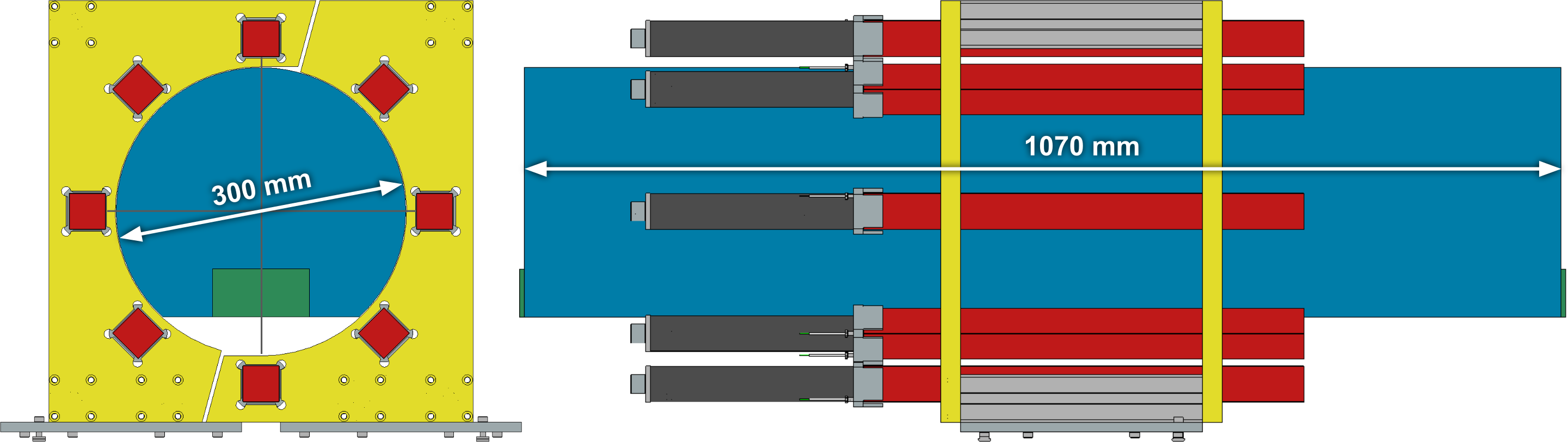}
    \caption{\it CAD back view (left) and side view (right) of the FLASHForward beam dump (blue) with the dump leakage calorimeter consisting of lead-glass bars (red) that are read out with Photo-Multiplier Tubes (PMTs) that are mounted with a light guide inside a plastic housing (dark gray). Two scintillating tiles (green) are taped to the front and back face of the dump. }
    \label{fig:cadFlashDump}
\end{figure}

The FLASHForward experiment at the FLASH facility~\cite{schreiber_freeelectron_2015, faatz_simultaneous_2016} at DESY typically operates bunch charges up to \SI{1}{\nano\coulomb}, corresponding to ${6.2 \times 10^9}$ electrons, at a rate of \SI{10}{\hertz}. The energy can be adjusted between \SI{400}{\mega\electronvolt} and \SI{1350}{\mega\electronvolt} with a relative full width at half maximum energy spread of $\lesssim 1\%$ with a linear chirp~\cite{darcy_flashforward_2019, lindstrom_emittance_2024}. These particle rates and energies are similar to those expected at the photon dump of the LUXE experiment~\cite{luxecollaboration_technical_2024}. Furthermore, the dump of the FLASHForward experiment is not obstructed by shielding and other elements, making it easily accessible for instrumentation.  

\begin{figure}
    \centering
    \includegraphics[width=\linewidth]{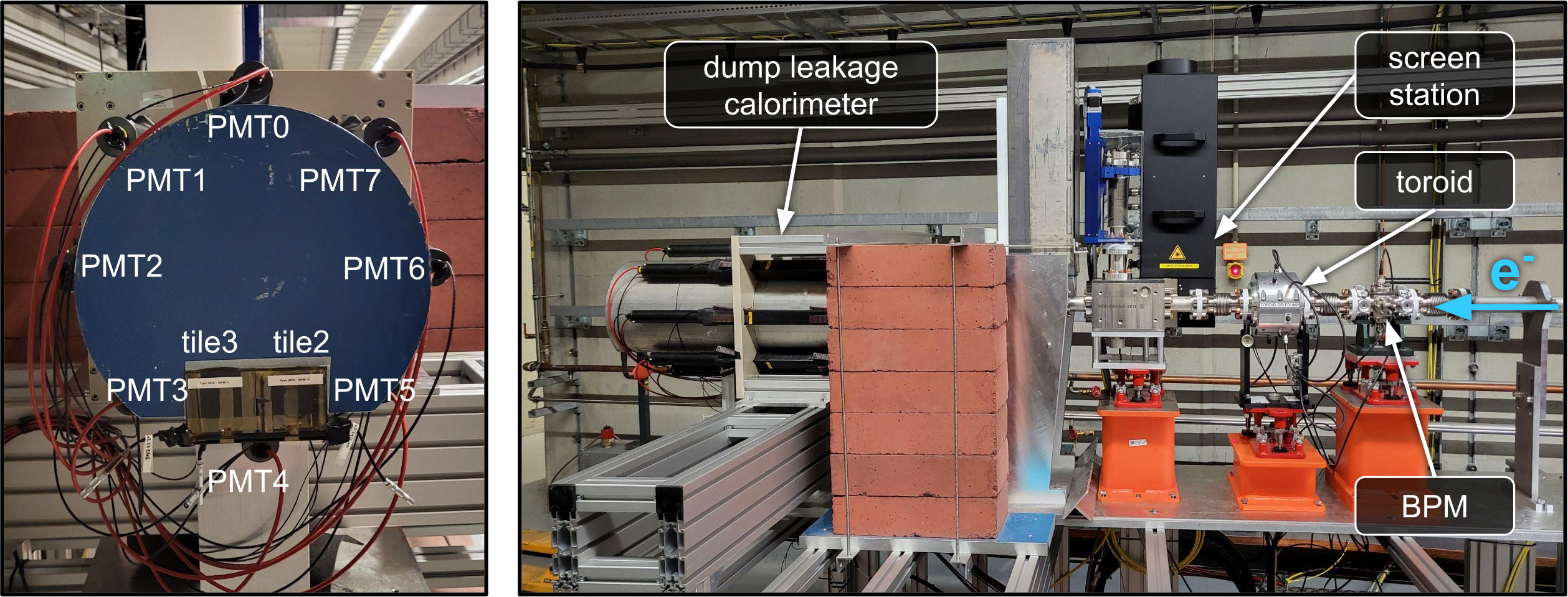}
    \caption{\it Photos of the dump section of the FLASHForward experiment at DESY. (left) View of the backside of the dump. The eight PMTs are mounted around the dump and labeled PMT0 to PMT7. Two scintillating tiles with SiPMs, labeled tile2 and tile3, are taped to the dump below the beam axis. (right) Side view of the end of the beamline where the electrons enter from the right side. They first pass a beam position monitor (BPM), a toroid beam charge transformer to measure the bunch charge, and a screen station with a retractable scintillating screen to measure the transverse beam profile before exiting the vacuum beam pipe and hitting the dump. }
    \label{fig:flashforwardPhoto}
\end{figure}

The CAD design and photos of the detector setup are shown in figures~\ref{fig:cadFlashDump} and \ref{fig:flashforwardPhoto}, respectively. The dump is made of an aluminum-magnesium alloy (AlMg). It has a cylindrical shape with a length of \SI{1070}{\milli\meter} and a diameter of \SI{300}{\milli\meter}. Assuming aluminum-like material properties, the dump length corresponds to approximately 12 radiation lengths. A segment on the bottom side of the dump is removed to allow for a secure mounting. The calorimeter consists of eight lead-glass bars, which are coupled to Photo-Multiplier Tubes (PMTs), mounted around the dump. The lead-glass bars have a cross section of \SI{38}{\milli\meter} $\times$ \SI{38}{\milli\meter}, a length of \SI{450}{\milli\meter}, and are of glass type TF1~\cite{lytkarinoopticalglassfactoryjsc_tf1_} as used in the GAMS-2000 spectrometer~\cite{binon_hodoscope_1986}. The PMTs are of type XP1911/UV from Photonis SAS~\cite{photoniss.a.s_photomultiplier_2007} and were used in the HERMES RICH detector~\cite{hermescollaboration_hermes_1999}. The lead-glass bars and the PMTs are connected via optical light guides made of plexiglas to transport the light. In addition to the lead-glass calorimeter modules, we mounted two scintillating tiles of size \SI{60}{\milli\meter} $\times$ \SI{60}{\milli\meter} $\times$ \SI{5}{\milli\meter} that are optically connected to silicon photomultipliers (SiPMs) via wavelength-shifting fibers before and after the dump below the beam axis. The PMTs and SiPMs were directly electronically connected to a CAEN V1730 waveform digitizer~\cite{caenspa_ds3153_2019} without additional preamplification. The digitizer was controlled via a modified version of the CAEN Wavedump software~\cite{dejong_caenv1730daq_2023}. The data acquisition was synchronized with the accelerator clock, using a delay to compensate for the position of the dump at the end of the facility. The digitizer acquired the full waveform of \SI{1}{\micro\second} - \SI{2}{\micro\second} for each channel and event with a \SI{122}{\micro\volt} resolution and a sampling interval of \SI{2}{\nano\second}. Approximately $10^3$ valid events per run were recorded.

In addition to the data from the dump leakage calorimeter, data from the accelerator beam diagnostics were recorded via the main accelerator control system~\cite{hensler_doocs_1996, karstensen_flashforward_2018}. Three devices were used at the end of the beamline before the dump, which are indicated in figure~\ref{fig:flashforwardPhoto}. A button-type beam position monitor (BPM) allows measurement of the transverse beam position with a position noise of about \SI{10}{\micro\meter}~\cite{treyer_design_2013, baboi_beam_2022}. A toroid beam charge transformer is used to precisely measure the bunch charge via passive induction and a resolution of about \SI{1}{\pico\coulomb}~\cite{werner_sensitivity_2011}. Finally, a screen station with a retractable scintillating screen that is read by a camera system enables the determination of the shape and position of the beam~\cite{wiebers_scintillating_2013}. The latter system has a resolution of \SI{10.8}{\micro\meter} in $x$ and \SI{11.4}{\micro\meter} in $y$. It was used during a single data-taking campaign to ascertain the size of the beam spot. We determined a full width at half maximum size on the screen of \SI{0.23(1)}{\milli\meter} in the horizontal direction and \SI{0.63(2)}{\milli\meter} in the vertical direction for a \SI{100}{\pico\coulomb} bunch charge. The screen was intermittently retracted during measurements, depending on the diagnostic requirements.

\section{Measurements}\label{sec:measurements}

The data presented here were obtained in several measurement campaigns. They are publicly available together with the analysis code~\cite{schulthess_ivoschulthess_2025, schulthess_raw_2025}. The accelerator electron energy was set to accommodate the requirements of the other users of the facility and could therefore not be specifically chosen or varied for this study. Error bars are included in all figures but are generally smaller than the marker size and therefore not visible. The signal uncertainty for each beam setting is taken as the statistical uncertainty of the mean, estimated as the standard deviation of the measured signals divided by the square root of the number of events.

The first successful campaign at FLASHForward in November 2023 allowed investigation of the position dependence of the signal described in section~\ref{sec:position}. The electron energy was \SI{1050}{\mega\electronvolt}. 

A later campaign in March 2024 with a beam energy of \SI{1020}{\mega\electronvolt} was used to find the observable with the most linear response, explained in section~\ref{sec:observables} and enabled us to perform the charge calibration described in section~\ref{sec:charge}. 

The last campaign in April 2024 with a beam energy of \SI{1200}{\mega\electronvolt} allowed comparison of the various sensor options described in section~\ref{sec:sensor}.
It was used to validate the charge calibration and to determine the precision and accuracy of the bunch charge estimates, which is described in detail in section~\ref{sec:charge}.

\subsection{Observables}\label{sec:observables}

The analysis of the waveform (i.e.\ the signal as a function of time) typically focuses on one of the following three observables: the peak amplitude, the numerical waveform integral, or the time-over-threshold. A baseline was determined from data acquired before the signal arrived and was subtracted from all signals. To evaluate the time-over-threshold, a fixed threshold of \SI{120}{\milli\volt} above the baseline was used to provide a consistent definition across all charge settings. 

The three observables show different dependences on the electron bunch charge because they probe different features of the waveform shape. At higher bunch charges, the PMT response becomes non-linear and the pulse height saturates, which causes the peak amplitude to flatten. The waveform integral is less affected because it also captures a broadening of the main signal peak. The time-over-threshold is defined with respect to a fixed threshold above the baseline and chosen such that it includes only the main peak. In the saturated regime, an additional signal therefore primarily increases the pulse width above threshold, as the pulse shape broadens rather than scaling purely in amplitude, leading to a steeper charge dependence. Using a relative threshold instead of an absolute one yields nearly identical results.

\begin{figure}
    \centering
    \includegraphics[width=0.7\linewidth]{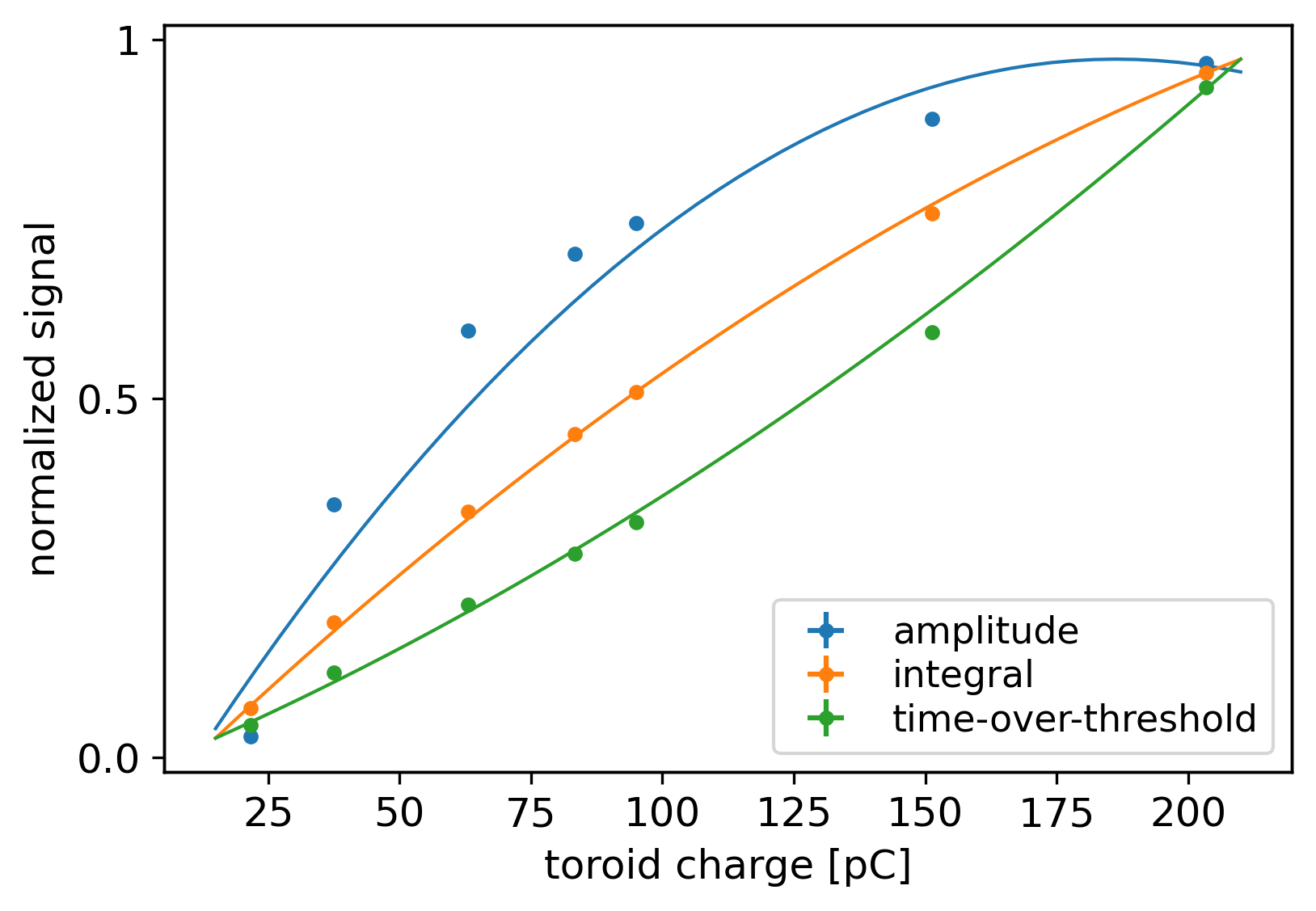}
    \caption{\it Normalized signals of the recorded waveforms for various electron bunch charges. The three analysis methods that are compared are the waveform's amplitude (blue), integral (orange), and time-over-threshold (green). The data are fitted with quadratic polynomials. }
    \label{fig:methodComparison}
\end{figure}

Figure~\ref{fig:methodComparison} shows the normalized observables of the three methods for comparison, together with the result of an orthogonal distance regression (ODR)~\cite{boggs_orthogonal_1989} of a quadratic polynomial.\footnote{To account for uncertainties in both the independent and the dependent variable, we employed orthogonal distance regression (ODR). The uncertainties of the resulting parameters differ significantly from those obtained with ordinary least squares (OLS).} The signals shown here correspond to the full calorimeter configuration described in section~\ref{sec:sensor}, using the average of PMT2 and PMT6. Given the evident non-linearity of the data and the absence of a physics-based model, we employed a quadratic function as the simplest non-linear approximation which empirically approximated the data. In our assessment, we used the waveform integral, which demonstrated better linearity, particularly in the higher charge range, along with a closer consistency with the quadratic model. As the model is purely empirical, statistical fit parameters, such as residual variance, are not interpreted as indicators of physical validity.

\subsection{Sensors}\label{sec:sensor}

One of the major limitations of the tested dump leakage calorimeter prototype is the non-linear signal response measured in the data-taking campaigns. In this first prototype, the signals of three sensor configurations were acquired. The first is the signal of the complete calorimeter, including the lead-glass bar, the light guide, and the PMT. In the second configuration, the lead glass was optically separated from the PMT with the light guide, such that the Cherenkov light created in the lead glass could not reach the PMT. In this case, the signal originates from charged particles of the electromagnetic shower traversing other components such as the light guide and the PMT entrance window directly, which can produce light via the Cherenkov or scintillation process. These additional light contributions are therefore expected to influence the signal response and are likely to contribute to the observed non-linearity by increasing the signal load on the PMT and readout chain, especially at higher bunch charges. Finally, the signals from the plastic scintillators with the SiPMs were recorded. 

\begin{figure}
    \centering
    \includegraphics[width=0.7\linewidth]{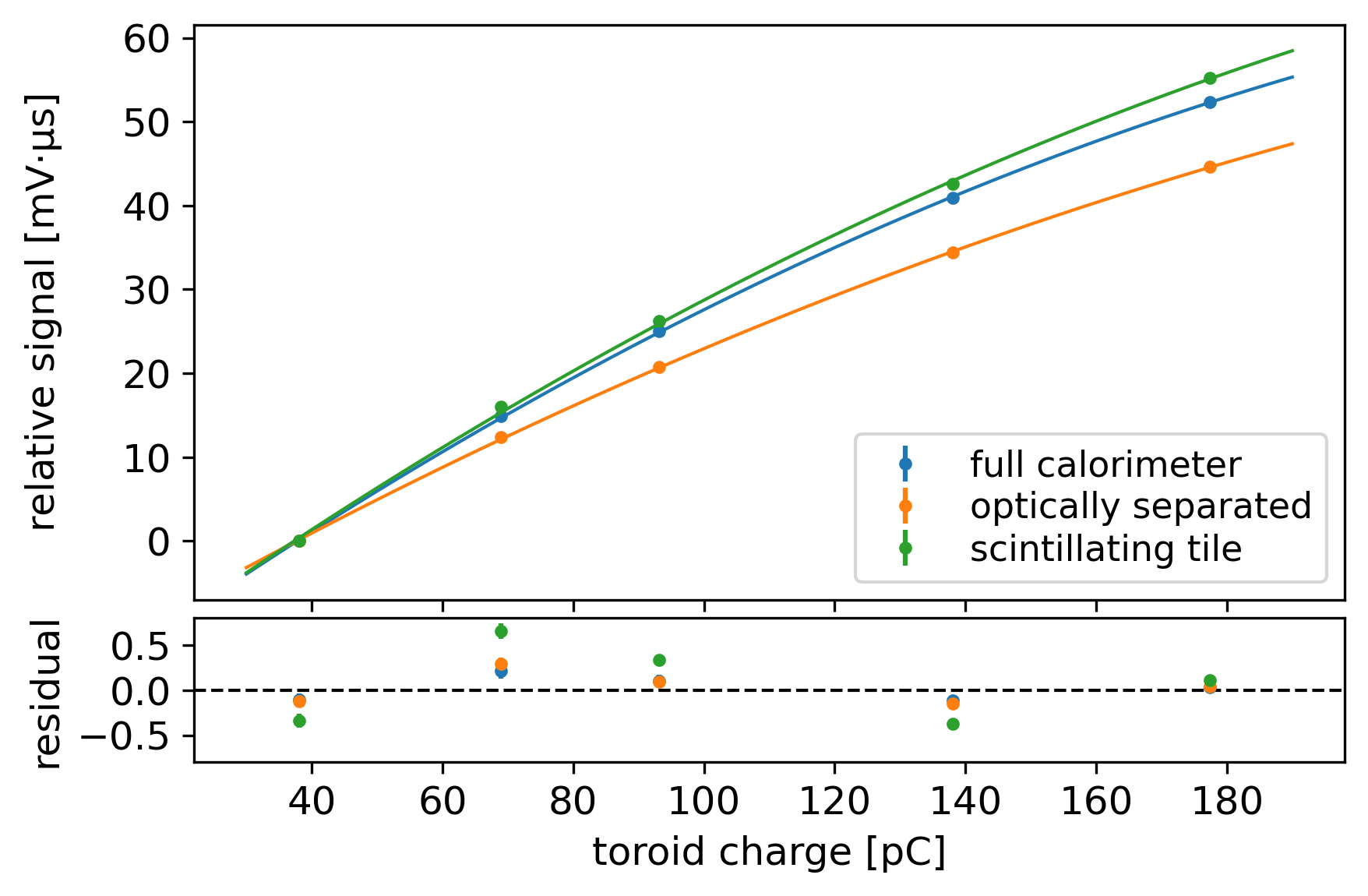}
    \caption{\it Relative signal response of various sensor settings as a function of the electron bunch charge. The three settings shown are the response of a full calorimeter module (blue), the optically separated configuration (orange), and a scintillating tile (green). The data are fitted with quadratic polynomials. Displayed below are the residuals, representing the difference between the observed data and the fitted values. }
    \label{fig:sensorComparison}
\end{figure}

The relative signal response in a charge scan is shown in figure~\ref{fig:sensorComparison}. All signals were shifted such that the value at the lowest bunch charge of \SI{38}{\pico\coulomb} is set to zero, i.e.\ the signal at this point is subtracted from all measurements for a relative comparison. The different sensor configurations show a similar response to increasing charges, that is, a flattening of the curve. The full calorimeter module configuration was used for the analysis as it was employed in all measurement campaigns.

\subsection{Position}\label{sec:position}

\begin{figure}
    \centering
    \includegraphics[width=0.7\linewidth]{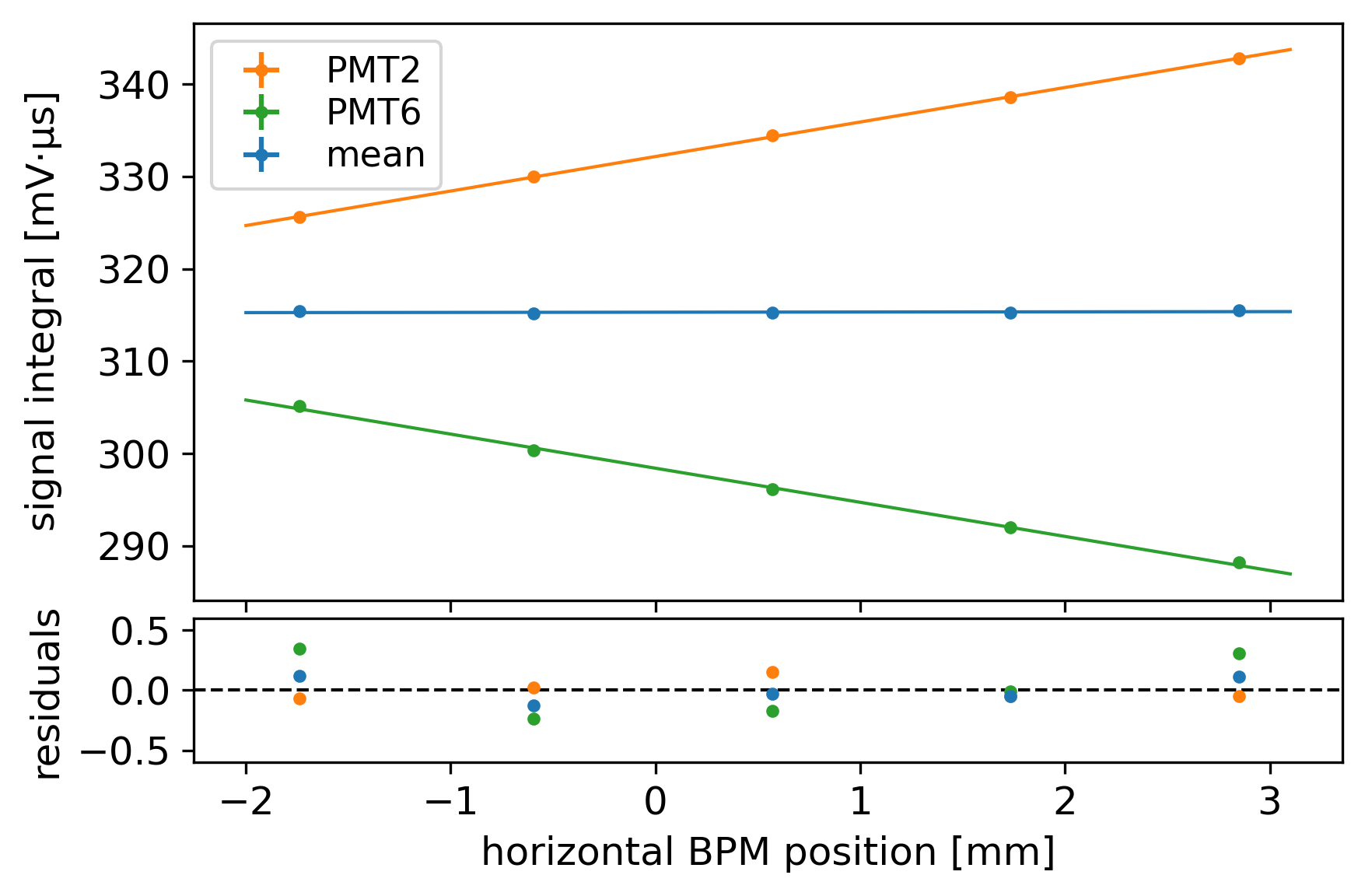}
    \caption{\it Signal integral of PMT2 (orange), PMT6 (green), and the mean of the two (blue) as a function of the horizontal position measured by the last BPM before the dump along with a linear fit. The residuals are shown below, representing the difference between the observed data and the fitted values. }
    \label{fig:positionScan}
\end{figure}

The transverse position of the electron bunch can be steered with beamline magnets which are roughly \SI{9}{\meter} upstream of the beam dump. A horizontal position scan was performed. The bunch charge for this measurement was \SI{99(1)}{\pico\coulomb}. The signal integral as a function of the horizontal position is presented in figure~\ref{fig:positionScan} for PMT2, PMT6, and their average. In principle, the beam position could be reconstructed using all calorimeter modules. However, due to the limited calibration data available for inter-calibrating all channels, only the opposing PMT pair (PMT2 and PMT6) was used, providing a robust and approximately geometry-symmetric measurement. An ODR of a straight line yields a slope of \SI{3.74(3)}{\milli\volt\micro\second\per\milli\meter} for PMT2, \SI{-3.69(9)}{\milli\volt\micro\second\per\milli\meter} for PMT6, and \SI{0.02(4)}{\milli\volt\micro\second\per\milli\meter} for the mean of the two. If the beam position was moved closer or farther from a calorimeter module, the signal increased or decreased, respectively. However, the mean of the two stayed constant over the tested range. The offset between the signal levels of PMT2 and PMT6 at a nominal BPM position of \SI{0}{\milli\meter} reflects differences in their absolute response and the fact that the BPM zero does not necessarily correspond to the geometric center of the dump.

This measurement shows two important characteristics. The setup has the capability of resolving the position of the beam. The precision is obtained by propagating the measured signal uncertainty and the parameters of the linear signal--position relation through Monte Carlo uncertainty propagation. In this procedure, fluctuations of the measured signal and variations of the fitted parameters within their uncertainties are translated into a distribution of reconstructed positions, from which the precision is extracted. This yields a precision of \SI{15}{\micro\meter} and \SI{39}{\micro\meter} for PMT2 and PMT6, respectively. While the two PMTs form a geometrically symmetric pair by design, the uncertainty of the fitted slope is significantly larger for PMT6, which directly propagates into a worse position precision. This increased uncertainty arises from the larger scatter of the PMT6 measurements around the fitted linear trend. Second, the average signal remains unaffected by the beam's position and is therefore insensitive to beam jitter. Thus, this average signal of the two channels (PMT2 and PMT6) is used for the charge analysis. This demonstrates that a position-independent observable can be constructed, which is advantageous for a robust charge measurement.

\subsection{Charge}\label{sec:charge}

Two separate measurement campaigns were conducted to investigate how effectively the dump leakage calorimeter can measure the bunch charge. The charge calibration was performed using data from the March 2024 campaign with an electron beam energy of \SI{1020}{\mega\electronvolt} and bunch charges between \SI{20}{\pico\coulomb} and \SI{200}{\pico\coulomb}, by relating the integral of the PMT signal to the bunch charge measured with the beamline toroid. The lower limit of the charge range is given by the working range of the BPM. At bunch charges above \SI{200}{\pico\coulomb}, the signal response of the dump leakage calorimeter becomes too flat to produce reliable measurements. This limitation arises from the non-linear response of the current prototype at high particle fluxes and can be mitigated in future iterations by optimizing the sensor configuration and operating parameters. Tests with lower PMT high-voltage settings reduced the overall signal amplitude but did not qualitatively change the observed saturation behaviour, indicating that the non-linearity is not solely determined by the PMT gain. The upper end of the tested bunch charge range is comparable to the particle fluxes expected at the photon beam dump of the LUXE experiment ($\sim 10^6 - 10^9$ photons per bunch), while the lower end is limited by machine constraints rather than detector performance. 

\begin{figure}
    \centering
    \includegraphics[width=0.7\linewidth]{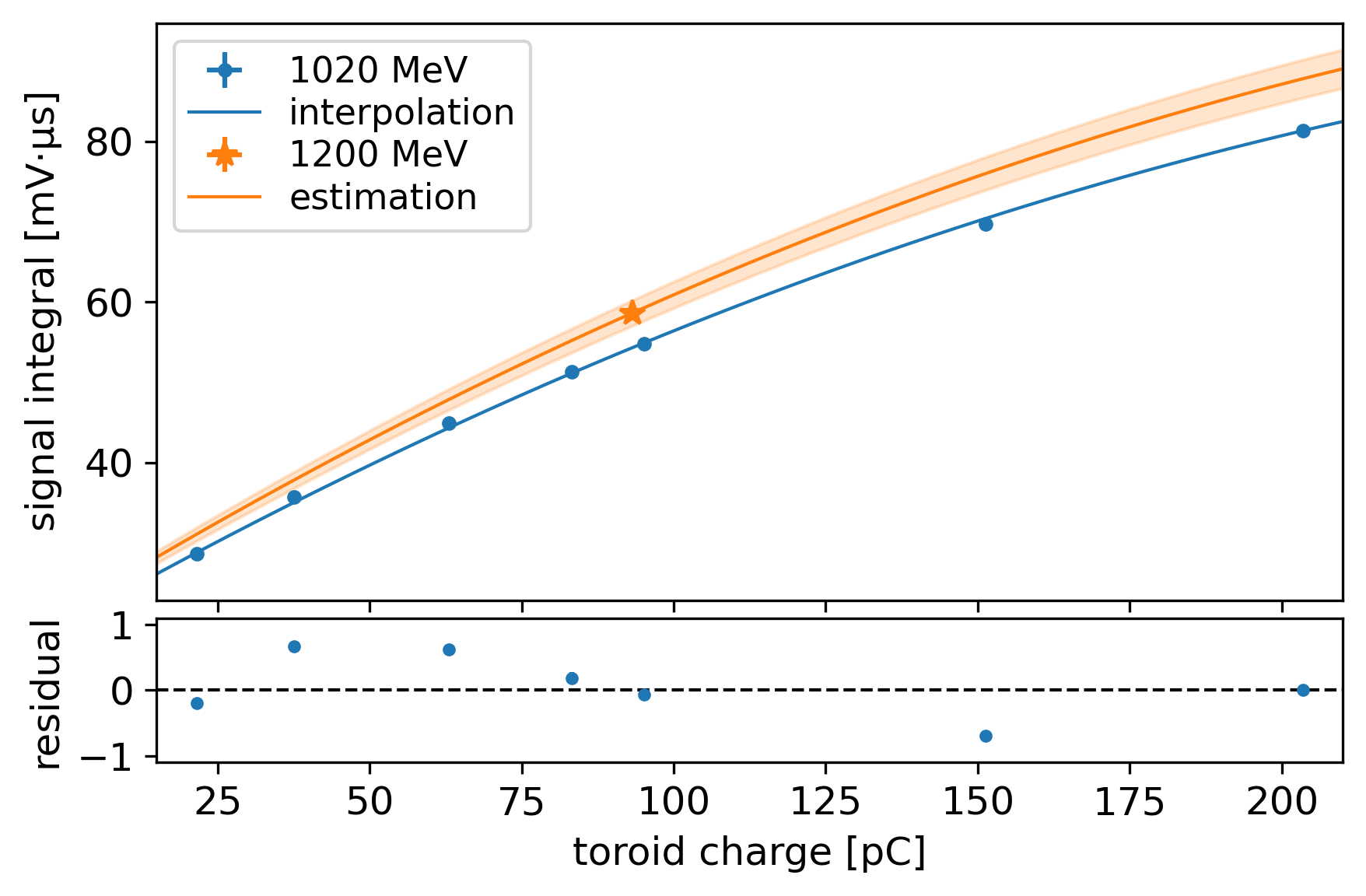}
    \caption{\it Charge scan of the dump leakage calorimeter. It shows the measured signal integral of the waveform as a function of the electron bunch charge. The calibration measurements (blue \ding{108}) are fitted with the quadratic function given in equation~(\ref{eq:chargeIntegral}) with the parameter $C_E$ fixed to one. One charge setting of the validation measurements (orange \ding{72}) has to be used to calibrate for the different electron beam energy. The estimation curve is shown in orange with its uncertainty band. The residuals are shown below, representing the difference between the observed data and the fitted values. }
    \label{fig:chargeScan}
\end{figure}

The calibration measurements are shown in blue in figure~\ref{fig:chargeScan}. The data are interpolated with a quadratic function given in equation~(\ref{eq:chargeIntegral})
\begin{equation}\label{eq:chargeIntegral}
    I = C_E \times (p_2 Q^2 + p_1 Q + p_0) \ ,
\end{equation}
where $C_E$ is a constant to account for the particle energy differences between the data-taking periods. For the calibration data set, it is fixed ${C_E = 1}$. 

\begin{figure}
    \centering
    \includegraphics[width=0.7\linewidth]{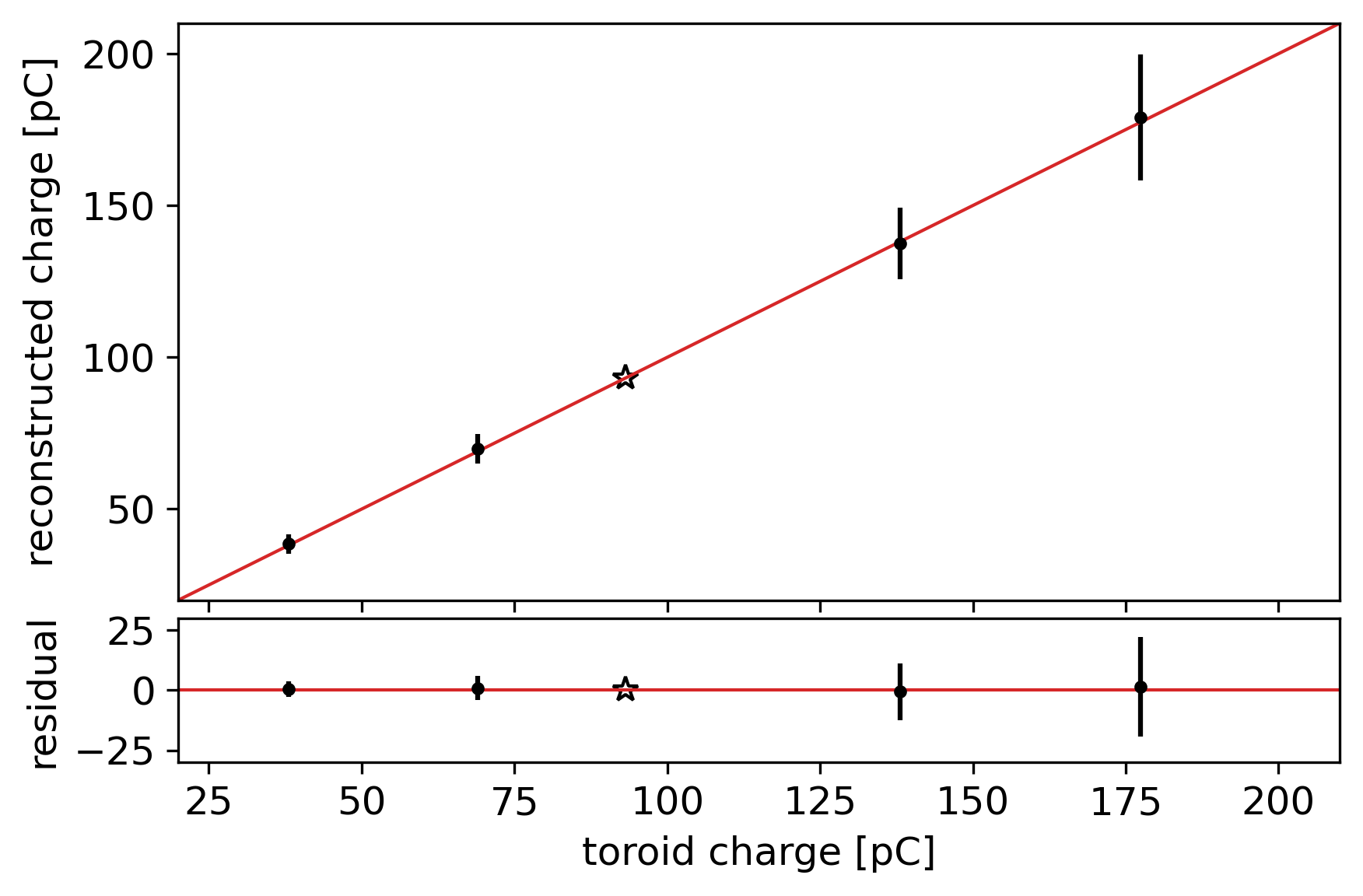}
    \caption{\it Reconstructed charge from the dump leakage calorimeter for one representative energy-calibration choice (\SI{93}{\pico\coulomb}, \ding{73}). The identity line is shown in red. The residuals are shown below, representing the difference between the observed data and the fitted values. }
    \label{fig:chargeEstimate}
\end{figure}

\begin{figure*}
    \centering
    \includegraphics[width=\linewidth]{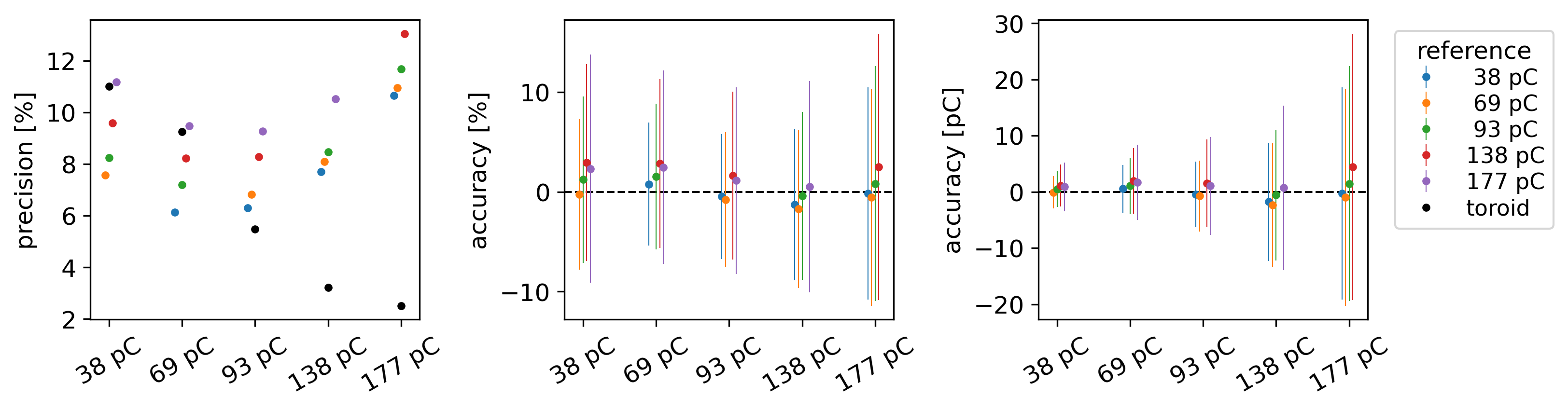}
    \caption{\it Results of the bunch charge estimates of the dump leakage calorimeter. It shows the relative precision (left), the relative accuracy (middle), and the absolute accuracy (right) of the charge estimation. The x-axis corresponds to the bunch charge used for the energy calibration, and the colored data points represent the reconstructed values for all other charge settings.  The black data represent the charge measurement of the toroid. }
    \label{fig:precisionAccuracy}
\end{figure*}

The second campaign served to validate the calibration. The electron energy was \SI{1200}{\mega\electronvolt}. Given the difference, we used the data point of one charge setting to fit the constant $C_E$ of equation~(\ref{eq:chargeIntegral}) while having the other parameters $p_i$ fixed at the calibration measurement values. Using a single charge setting for the energy calibration preserves the remaining measurements as an independent validation of the charge reconstruction. This is indicated in figure~\ref{fig:chargeScan} by the charge measurement at \SI{93}{\pico\coulomb} and the estimation curve in orange. The comparison of the toroid charge measurement with the estimated charge of the dump leakage calorimeter $Q$ and its uncertainty $\sigma_Q$ is presented in figure~\ref{fig:chargeEstimate}. It shows that both measurements are in good agreement. Figure~\ref{fig:chargeEstimate} illustrates this comparison for one representative choice of energy calibration, while a quantitative assessment of the linearity and accuracy for all possible calibration choices is presented in figure~\ref{fig:precisionAccuracy}. The consistency of the results for all calibration choices indicates that the scaling with $C_E$ in equation~(\ref{eq:chargeIntegral}) is adequate within the quoted uncertainties. The uncertainties of the energy calibration and the charge were estimated using Monte Carlo uncertainty propagation, in which the input parameters of the calibration function and the measured signal are varied within their uncertainties and propagated to the reconstructed charge. 

To assess systematic effects, the value of the energy calibration $C_E$ is obtained for each measured charge setting and the remaining charges were reconstructed. The results are summarized in figure~\ref{fig:precisionAccuracy}. The left subfigure shows the relative precision ${\sigma_Q / Q}$, while the middle and right subfigures show the relative accuracy ${(Q - Q_t) / Q_t}$ and absolute accuracy ${Q - Q_t}$, respectively. $Q_t$ is the charge measured by the toroid, which has a mean precision of \SI{5}{pC} across all measurements, corresponding to a relative precision that varies with bunch charge. This exceeds the intrinsic resolution of the toroid ($\sim \SI{1}{pC}$, see section~\ref{sec:setup}) and results from varying the bunch charge using a scraper, which likely introduces additional jitter through orbit fluctuations~\cite{schroder_tunable_2020}. The results in figure~\ref{fig:precisionAccuracy} show that the typical precision is on the order of 10\%, while the accuracy is at the few-percent level across the different choices of energy calibration. The x-axis corresponds to the bunch charge at which the energy calibration is performed, and the precision and accuracy shown reflect the reconstruction of all other charge settings. Consequently, the precision degrades when calibration points at the edges of the charge range are used, in particular at high charge where the uncertainty of the energy calibration becomes the largest. The results also indicate systematic discrepancies in the accuracy that originate from the non-quadratic behavior of the calorimeter response, assumed in equation~(\ref{eq:chargeIntegral}), which is already visible in the calibration fit shown in figure~\ref{fig:chargeScan}. The error bars in figures~\ref{fig:chargeEstimate} and \ref{fig:precisionAccuracy} reflect the total propagated uncertainty, which is dominated by the uncertainties of the calibration parameters and increases towards higher charges, where the detector response becomes flatter and more sensitive to the calibration. Fixing these parameters would significantly reduce the uncertainty. The visible scatter of the data points is smaller than the error bars, as systematic effects do not induce point-to-point fluctuations.

\section{Shower leakage simulations for electron and photon beams}\label{sec:simulations}

Even though the prototype of the dump leakage calorimeter was commissioned and tested at an electron beam dump, its main goal is to measure the flux of a pulsed photon beam. The simulations presented here are used to compare the development of electromagnetic showers initiated by electrons and photons in the dump geometry and to justify the use of electron-beam measurements as a proxy for photon-beam operation, an approach that is also employed in other experiments where electron-based calibrations are transferred to photons with dedicated corrections~\cite{ATLAS:2023mnw}. Our simulations focus on the shower leakage into the calorimeter rather than on the absolute detector response. A detailed detector-response simulation, including optical photon production and sensor response, is beyond the scope of this proof-of-concept study. Accordingly, the simulation is restricted to energy deposition as the observable directly relevant to the measured calorimeter signal, which is expected to scale with the detector response in the regime considered. 

To assess the applicability of this approach to photon-beam operation, we performed simulations using the G4beamline tool v3.08~\cite{roberts_g4beamline_2022}, which is based on Geant4~\cite{agostinelli_geant4a_2003, allison_geant4_2006, allison_recent_2016}, with the QGSP\_BERT physics list and its default settings. We compare the development of the electromagnetic shower and the energy deposited in the lead glass, i.e., the sum of the energy lost by all particle tracks that pass through the lead glass. The cases in which electrons or photons are disposed of in the dump were compared for a bunch with $6.24 \times 10^8$ particles, corresponding to an electron bunch with a charge of \SI{100}{\pico\coulomb}. The beam energy was varied between \SI{300}{\mega\electronvolt} and \SI{3}{\giga\electronvolt} and for each particle type and energy, 25 trials with $10^5$ particles were simulated.

\begin{figure}
    \centering
    \includegraphics[width=0.7\linewidth]{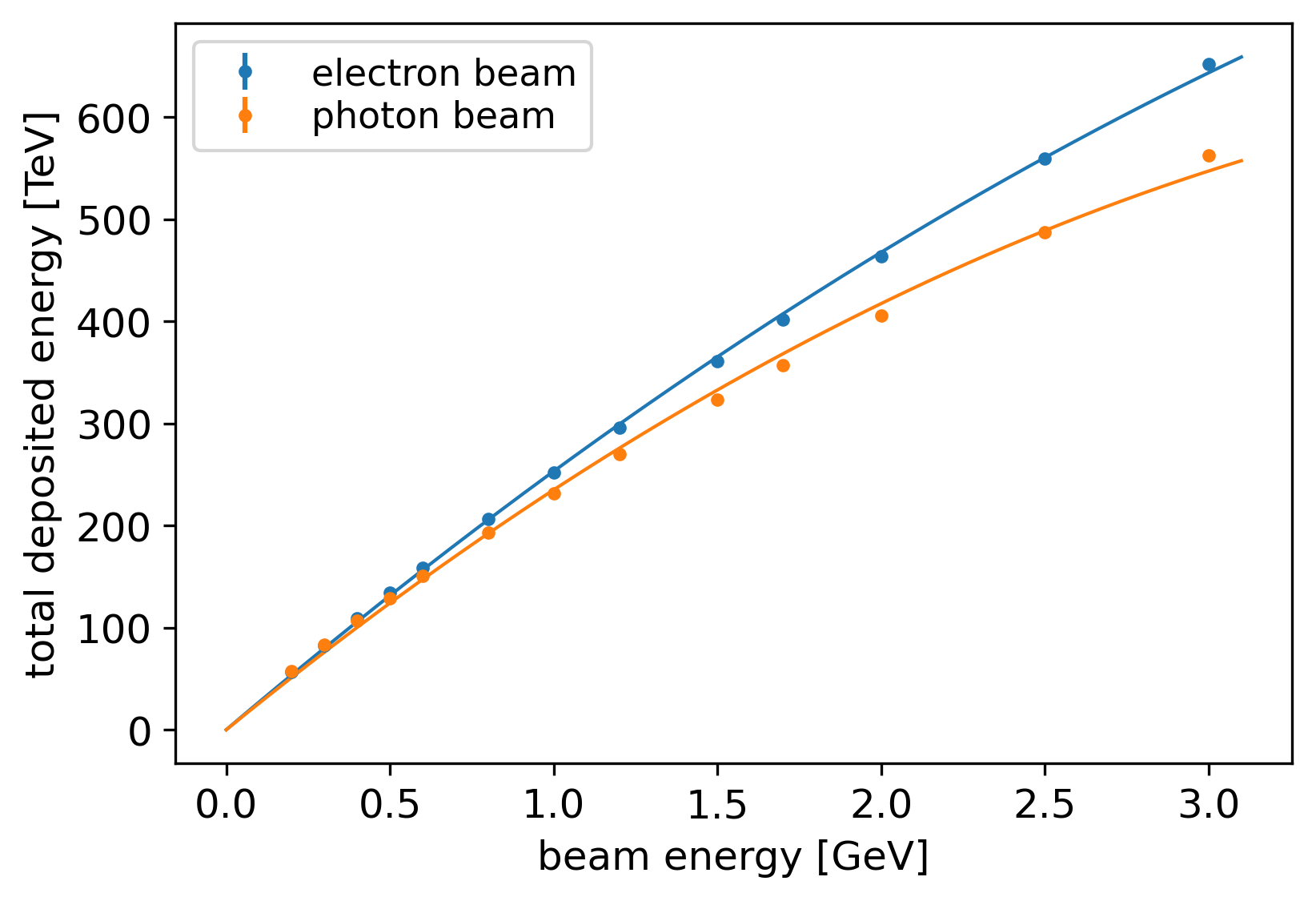}
    \caption{\it Simulations of the deposited energy in a lead-glass bar of the calorimeter module as a function of the initial particle's energy for electrons (blue) and photons (orange), shown for a bunch of $6.24 \times 10^8$ particles (\SI{100}{\pico\coulomb}). The statistical uncertainty of the simulation of less than 3\% is smaller than the size of the data point. The solid lines are ordinary least-squares fits and serve only as a guide to the eye. }
    \label{fig:electronPhotonDiff}
\end{figure}

The results of the simulation are presented in figure~\ref{fig:electronPhotonDiff}. At energies below \SI{500}{\mega\electronvolt}, the deposited energy for electrons and photons is the same within the statistical uncertainty of the simulation. At higher energy, both start to level off at slightly different rates, which leads to a lower energy deposition for a photon beam. This can be explained by the slight difference in the longitudinal shower profile of the two species and the geometrical acceptance of the lead glass. Nevertheless, the discrepancy only requires a recalibration for photons and does not pose an issue for the photon flux measurements in future experiments or facilities.

\section{Conclusion}\label{sec:conclusion}

We presented a novel dump leakage calorimeter capable of operating in the high-flux, high-energy regime where diagnostics are challenging. It enables measurement of the electron or photon flux without affecting the beam before entering the dump. We demonstrated how a combination of detector channels located at different positions of the dump can be used to eliminate the position dependence and that we can resolve the position of the beam with a precision on the order of tens of micrometers. Furthermore, we showed that it can measure the number of electrons in a bunch with a typical precision on the order of 10\% and an accuracy at the few-percent level, and that the concept is also applicable for high-energy photons. However, non-linearities of the calorimeter's response lead to systematic deviations of the bunch charge estimate that should be mitigated in future upgrades of the detector.

\section{Outlook}\label{sec:outlook}

To mitigate the non-linear signal response of the dump leakage calorimeter and to further improve of the setup, several aspects can be addressed in a future upgrade: 

\begin{itemize}
    \item The position and orientation of the calorimeter modules were not fully optimized in this first prototype. Simulations can be performed to optimize the signal in the calorimeter so that shower leakage can be measured with the best sensitivity in the charge range of interest. 
    \item Preliminary simulations and laboratory tests indicate that much of the signal in the calorimeter modules originates from light produced in other components rather than the lead glass itself. The quenching of the scintillation light yield~\cite{birks_scintillations_1951} together with a saturation of the PMT~\cite{hamamatsuphotonicsk.k._photomultiplier_2017} are likely causes of the non-linear signal response at higher bunch charges. This could be mitigated by removing the light guide or by optimizing the positioning of the calorimeter modules so that the particle flux through the light guide is reduced. 
    \item Radiation damage must be considered, especially in the case of the LUXE dump leakage calorimeter. The lead glass used for this prototype cannot handle the radiation doses that are expected at LUXE in such a configuration. The more radiation-hard lead glass TF101 could be used. Details must be discussed when the specific detector layout is worked out. An LED pulser is being developed to calibrate the calorimeter and monitor the darkening of the lead-glass bars. 
    \item In a future experiment like LUXE, the photon spectrum will not be monochromatic but will span an energy range of several GeV. This will complicate the measurement and require additional dedicated studies and tests. 
\end{itemize}

\acknowledgments{
The authors appreciate the technical support of Karsten Gadow, Sven Karstensen, Michelle Klotz, Beata Liss, Kai Ludwig, Frank Marutzky, Richie Nölling, Amir Rahali, Vladimir Rybnikov, and Andrej Schleiermacher. The experiment was performed at the beam-driven plasma-wakefield experiment FLASHForward at the FLASH facility of the Deutsches Elektronen-Synchrotron DESY in Hamburg, Germany. The AI-assisted tools OpenAI ChatGPT, Microsoft Copilot, and Writefull were used to help phrasing the text of this manuscript. This work was supported by the Swiss National Science Foundation under grants no. 214492 and 230596, and by Helmholtz ARD, the Helmholtz IuVF ZT-0009 program. }

\section*{Conflict of interest}
The authors have no conflict of interest to disclose.

\section*{Author contributions}

Contribution of all authors according to ANSI/NISO~\cite{nisocreditworkinggroup_ansi_2022}. \\

\textbf{A.~Athanassiadis:} data curation (supporting); investigation (equal); methodology (equal); software (equal); validation (equal); writing – review \& editing (supporting). 
\textbf{T.~Behnke:} writing – review \& editing (supporting). 
\textbf{J.~Björklund Svensson:} investigation (supporting); writing – review \& editing (supporting). 
\textbf{O.~Borysov:} conceptualization (equal); writing – review \& editing (supporting). 
\textbf{M.~Borysova:} conceptualization (equal); writing – review \& editing (supporting). 
\textbf{L.~Boulton:} investigation (supporting); writing – review \& editing (supporting). 
\textbf{S.~Chindaratchakul:} formal analysis (supporting); investigation (supporting); software (supporting); visualization (supporting); writing – review \& editing (supporting). 
\textbf{B.~Heinemann:} writing – review \& editing (supporting).
\textbf{L.~Helary:} conceptualization (supporting); data curation (supporting); formal analysis (supporting); funding acquisition (equal); investigation (equal); methodology (equal); project administration (equal); software (equal); supervision (equal); validation (equal); writing – review \& editing (equal). 
\textbf{R.~Jacobs:} investigation (supporting); writing – review \& editing (supporting). 
\textbf{A.~Kanekar:} investigation (supporting); writing – review \& editing (supporting). 
\textbf{J.~List:} writing – review \& editing (supporting). 
\textbf{T.~Long:} investigation (supporting); writing – review \& editing (supporting). 
\textbf{T.~Marsault:} investigation (equal); writing – review \& editing (supporting). 
\textbf{F.~Peña:} investigation (supporting); writing – review \& editing (supporting). 
\textbf{S.~Schmitt:} resources (equal); writing – review \& editing (supporting). 
\textbf{S.~Schröder:} investigation (supporting); writing – review \& editing (supporting). 
\textbf{I.~Schulthess:} data curation (lead); formal analysis (lead); funding acquisition (equal); investigation (equal); methodology (lead); project administration (equal); software (equal); supervision (equal); validation (equal); visualization (lead); writing - original draft (lead); writing – review \& editing (equal). 
\textbf{S.~Wesch:} investigation (supporting); writing – review \& editing (supporting). 
\textbf{M.~Wing:} writing – review \& editing (supporting). 
\textbf{J.~Wood:} investigation (supporting); writing – review \& editing (supporting).

\section*{Data availability}
The data and analysis that support the findings of this study are openly available~\cite{schulthess_ivoschulthess_2025, schulthess_raw_2025}.

\bibliographystyle{JHEP}
\bibliography{ref.bib}

\end{document}